\documentclass[conference]{IEEEtran}
\IEEEoverridecommandlockouts
\usepackage{cite}
\usepackage{amsmath,amssymb,amsfonts}
\usepackage{algorithm}
\usepackage{algorithmicx}
\usepackage{graphicx}
\usepackage{textcomp}
\usepackage{xcolor}
\usepackage{xcolor}
\usepackage[utf8]{inputenc}
\usepackage{amsthm}
\usepackage{graphicx}
\usepackage{algpseudocode}
\usepackage{amsfonts,amsmath,amssymb,stmaryrd }
\usepackage{amsthm}
\usepackage{tikz,lipsum,lmodern}
\usepackage[most]{tcolorbox}
\usepackage{listings}
\usepackage{amsmath}  
\usepackage{amssymb} 
\usepackage{newclude} 

\usepackage{url}
\usepackage{graphicx,color}
\usepackage{booktabs}
\usepackage{listings}
\usepackage{color}

\usepackage{pdfpages} 

\definecolor{lightgray}{rgb}{.9,.9,.9}
\definecolor{darkgray}{rgb}{.4,.4,.4}
\definecolor{purple}{rgb}{0.65, 0.12, 0.82}

\lstdefinelanguage{script}{
  keywords={},
  keywordstyle=\color{blue}\bfseries,
  ndkeywordstyle=\color{darkgray}\bfseries,
  identifierstyle=\color{black},
  commentstyle=\color{purple}\ttfamily,
  stringstyle=\color{red}\ttfamily,
  sensitive=true,
  breaklines=true,
  escapechar=\$
}
\lstdefinelanguage{javascript}{
  keywords={do, if, in, for, let, new, try, var, case, else, enum, eval, null, this, true, void, with, await, break, catch, class, const, false, super, throw, while, yield, delete, export, import, public, return, static, switch, typeof, default, extends, finally, package, private, continue, debugger, function, arguments, interface, protected, implements, instanceof},
  keywordstyle=\color{blue}\bfseries,
  ndkeywords={class, export, boolean, throw, implements, import, this},
  ndkeywordstyle=\color{darkgray}\bfseries,
  identifierstyle=\color{black},
  sensitive=false,
  comment=[l]{//},
  morecomment=[s]{/*}{*/},
  commentstyle=\color{purple}\ttfamily,
  stringstyle=\color{red}\ttfamily,
  morestring=[b]',
  morestring=[b]",
  breaklines=true,
  escapechar=\$
}

\begin{document}
\bstctlcite{IEEEexample:BSTcontrol} 

\title{
A Privacy-Preserving Platform for Recording COVID-19 Vaccine Passports\\
{}
\thanks{Identify applicable funding agency here. If none, delete this.}
}

\author{\IEEEauthorblockN{Masoud Barati, William J. Buchanan, Owen Lo}
\IEEEauthorblockA{\textit{Blockpass ID Lab} \\
\textit{Edinburgh Napier University}\\
Edinburgh, UK \\
\{M.Barati, B.Buchanan, O.Lo\}@napier.ac.uk}
\and
\IEEEauthorblockN{Omer Rana}
\IEEEauthorblockA{\textit{School of Computer Science \& Informatics} \\
\textit{Cardiff University}\\
Cardiff, UK \\
ranaof@cardiff.ac.uk}
}

\maketitle

\begin{abstract}

Digital vaccine passports are one of the main solutions which would allow the restart of travel in a post COVID-19 world. Trust, scalability and security are all key challenges one must overcome in implementing a vaccine passport.  Initial approaches attempt  to  solve  this  problem  by  using  centralised systems with trusted authorities.  However, sharing vaccine passport data between different organisations, regions and countries has become a major challenge. This paper designs a new platform architecture for creating, storing and verifying digital COVID-19 vaccine certifications. The platform makes use of InterPlanetary File System (IPFS) to guarantee there is no single point of failure and allow data to be securely distributed globally. Blockchain and smart contracts are also integrated into the platform to define policies and log access rights to vaccine passport data while ensuring all actions are audited and verifiably immutable. Our proposed platform 
realises General Data Protection Regulation (GDPR) requirements in terms of user consent, data encryption, data erasure and accountability obligations. We assess the scalability and performance of the platform using IPFS and Blockchain test networks.  

\end{abstract}


\section{Introduction}
COVID-19 has provided the world with an exceptional challenge. The first relates to the methods that countries use to suppress the spread of the virus and the second is how to re-start economies \cite{chen2020covid}. One of the core methods that could be used to re-start travel within each country is to implement a vaccine passport, and which could be used in future pandemics. Some of the initial approaches to the creation of a vaccine passport have turned to mandatory and centralised approaches using the PKI (Public Key Infrastructure) \cite{dgc}, others have proposed a non-mandatory and decentralised approach \cite{ouellette2021decentralized}. Within a centralised approach, we normally use a single trust authority to check the signatures on passports. This authority will then check the public key of the signer of the passport, and accept it, if it is a trusted entity. In this way, trusted health authorities can sign their own passports whenever someone is immunised. This approach, though, while relatively easy to implement, leaves the centralised infrastructure open to breaches, along with a major problem around the revocation of signing keys. 

Blockchain and smart contracts have been combined with platforms creating vaccine certificates, to increase trust in such certificates and privacy of citizens data.  In~\cite{HSJAY:2020}, a Blockchain-based solution involving self-sovereign identification, decentralised storage and re-encryption proxies was proposed. The approach used Ethereum smart contracts to run transaction calls and record events containing medical information and COVID-19 test updates. A Blockchain-based approach that prevents information tampering such as COVID-19 test results was presented in~\cite{MKSS:2020}. The approach supports monitoring of COVID-19 spread at an earlier stage via a passengers' travel history. A scalable, Blockchain-based platform for data sharing of COVID-19 and  vaccination passports was proposed in~\cite{HKGMK:2021}. The evaluation of the designed platform was performed  with 27 Blockchain nodes, each of which represents a European member state. A platform for secure COVID-19 passports and digital health certificate, called  NovidChain, was built in~\cite{ACKJ:2021}.  The platform restricted the propagation of COVID-19 while supporting privacy concerns and ensuring citizen's self-sovereignty for accessing their data. Although all these approaches provide secure mechanisms for creating vaccine certificates, none of them considers user consent before processing their personal data. Moreover, the data accountability of actors and the right to be forgotten which are the main requirements of General Data Protection Regulation (GDPR) were not studied in the reviewed approaches. GDPR is a European legislation for protecting personal data and enables citizens to control how their data is collected and manipulated by processing entities (actors)~\cite{Virvou:2017,ABRP:2021}.

In order to address the aforementioned challenges, this paper presents an architecture for a Blockchain-based platform that creates and verifies digital COVID-19 vaccine passports. The platform makes use of IPFS for storing and distributing citizens' passport data securely. It also involves a smart contract factory to improve transparency, trust and data privacy of citizens. 


Our proposed platform supports GDPR by implementing smart contracts that: (i) receive and record user consent, (ii) verify the operations of actors on passport data, and (iii) track  the realization of the right to be forgotten. The proposed smart contracts are deployed and tested using the Ethereum virtual machine, and their costs and mining time are investigated in Blockchain test networks.     

The rest of the paper is structured as follows. Section~\ref{back} briefly describes digital vaccine passports and IPFS. Section~\ref{sec:arch} describes the architecture of the platform together along with a number of protocols used in the implementation. Section~\ref{expr} assesses the time taken for creation of vaccine passports via IPFS and the costs required for deploying our implemented smart contracts and their transactions. Finally, Section~\ref{conclu} concludes the paper and identifies directions for future work. 


\begin{center}
\begin{figure*}
\begin{center}
\label{glass01}
\caption{Overview of the aims of GLASS\cite{glass2021}}
\end{center}
\end{figure*}
\end{center}

\section{Background \& Context}
\label{back}
This section provides brief descriptions of digital vaccine passports and the InterPlanetary File System (IPFS).

\subsection{Vaccine Passports}
 Phelan~\cite{phelan2020covid} outlines that those who hold vaccination passports could be exempt from self-isolation when they travel. 

\subsubsection{Digital Green Certificate}
In order to re-start international travel, the EU Commission has defined the Digital Green Certificate (DGC) programme \cite{dgc}, which defines three certificates: \emph{vaccination certificates}, \emph{test certificates} (NAAT/RT-PCR test or a rapid antigen test), and \emph{certificates for persons who have recovered from COVID-19}. This makes use of a centralised system that will receive the digitally signed records. Each of these will be signed by a trusted health care entity, and checked against the public key of that trusted entity. In the best case, key pairs would be issued to every trusted health care professional to sign passports. This would allow fine-grain control on signing and audit each signing authority. If there was a breach of the private key though, the public key would be revoked, and it would have a minimal impact. What is likely to happen is that a countrywide health authority will have a single signing key pair. A single breach of the private key will bring down every single DGC signed by that authority, as all of the passports will be marked as untrusted. For cybercriminals, the private key will be a key target, as it will be of significant worth to them on the open market.


\subsection{IPFS}
IPFS \cite{benet2014ipfs, ipfs} is a distributed, Peer-to-Peer (P2P) content sharing protocol. 
Traditionally, resources have been shared using a location-based approach (e.g. a URL or file path). In IPFS, every individual resource which is to be shared on the P2P network is identified and located using an identifier which is derived directly from the content of the resource.

IPFS does not centralise the storage of resources. Instead, peers on the network will distribute data individually and any peer who downloads the content will also become a distributor of that data.  Furthermore, the P2P nature of IPFS means we can reduce the latency involved in sourcing data, along with building resilience. IPFS can be used to represent any number of digital content including websites, folders, images, documents and even databases. 

IPFS distributes data on the P2P network by first breaking down resources into blocks of $256 \times 1024$ (262,144) bytes by default \cite{IPFS2021a}. Breaking down resources into blocks allows for deduplication (thus saving space) and storage of content in a distributed manner. Each block of data is content addressable using a Content-Identifier (CID). A Merkle Directed Acyclic Graph (DAG) data structure is used to represent each block of data and dictate the relationship between resources (e.g. parent folder and child files). Lastly, a Distributed Hash-Table (DHT) is used to allow peers to route and locate desired resources on the network (i.e. which peer is storing certain blocks and where they are located). The sections which follow describe CIDs, Merkle DAG and DHT in greater detail.

\noindent {\bf CID:} are used in IPFS to achieve the goal of sharing resources using a content-based approach, assigning a unique identifier to each content resource (e.g. text file, image file, images and so on) to be shared on the IPFS network. \emph{self-describing} identifier \cite{Multiformats2020} which uses a cryptographic hash (SHA-256) to address the content. 
Since the CID hash is derived directly from the content (i.e. the text "hello world"), the CID will remain the same regardless of the filename or any other associated metadata. This allows for a degree of assurance that one is downloading the correct content from the IPFS network so long as the CID is known and trusted. 

\noindent {\bf Merkle DAG}~\cite{merkle1987digital} is a cryptographic hashing function to represent and derive nodes of data from a root node. It is commonly described as a Merkle \emph{tree} due to the fact that the data structure it represents resembles an upside down tree (where the top node is the root). A Merkle tree can be used to verify the integrity of a data structure in a scalable manner since the root node's cryptographic hash can be used to verify the entire data structure represented by this algorithm. A Merkle DAG is an acyclic variation of a Merkle Tree with unique properties. Firstly, data structures represented by Merkle DAG are \emph{directed} which means there is a forward direction defining the relationship between two nodes (e.g. parent folder points to child document). 
In the context of IPFS, Merkle DAG is used to represent folders and files shared on the P2P network. 

\noindent{\bf DHT:} is a key-value lookup table which maps content hash values (CID) to the location of content (i.e. peers which host the files) in IPFS. DHT is used for routing and informing peers where resources are located on the P2P network \cite{benet2014ipfs}. Each peer on the IPFS network will store and maintain a list of known peers as new nodes join the network. The DHT algorithm implemented by IPFS is named Kademlia \cite{maymounkov2002kademlia}. Kademlia uses a unique address in the range of $0$ to $2^{256-1}$ to identify each peer on the IPFS network \cite{IPFS2021} and uses the exclusive OR ($XOR$) function to calculate the distance between each peer (thus allowing nodes to determine their nearest neighbours). This approach allows for the node lookup time  to be $O(log(N))$ (logarithmic time) \cite{IPFS2021, maymounkov2002kademlia} therefore ensuring scalability in the IPFS network even with a large number of peers.

\section{Systems Architecture}
\label{sec:arch}

\begin{figure}[t!]
\center
\includegraphics[width=0.65\columnwidth]{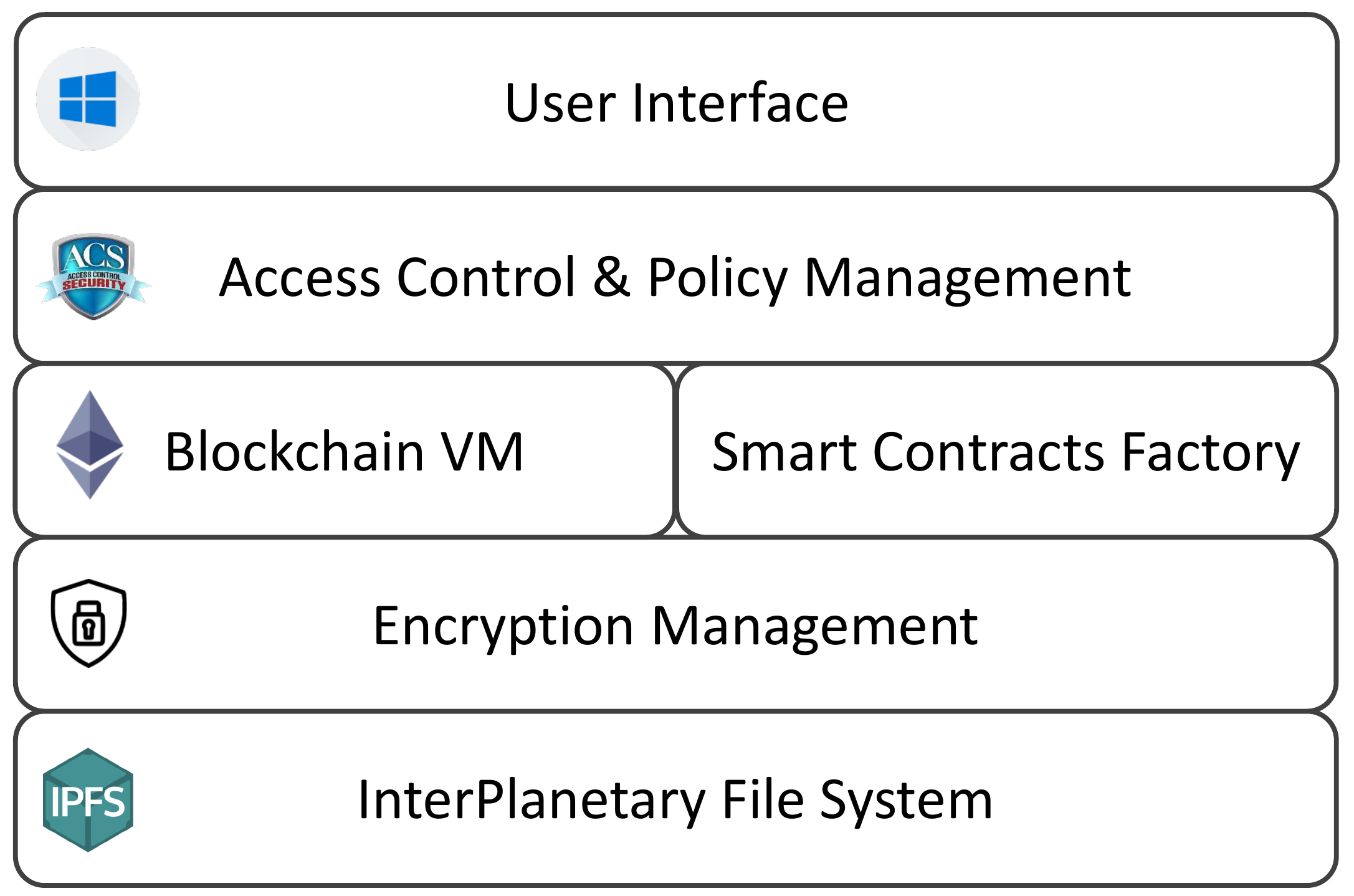}
\caption{The architecture of platform registering vaccine passports}
\label{fig:arc}
\end{figure}

A conceptual architecture for recording and verifying online COVID-19 vaccine passports is proposed in Fig.~\ref{fig:arc}, and consists of the following layers: \textbf{InterPlanetary file system (IPFS)} is a peer-to-peer network for storing and sharing the information relevant to the citizens who have been vaccinated. In Scotland, such information includes \textit{surname(s)}, \textit{forename(s)}, \textit{DOB}, \textit{country of vaccination}, \textit{identification number}, \textit{dose number}, \textit{dates of dosage}, \textit{manufacturer}, \textit{vaccine product} and \textit{vaccine/prophylaxis}. The information is stored in IPFS through an administrator in a medical centre offering the vaccines. IPFS generates a content identifier (CID), which is a label used to point to each citizen's record/ file.       
\\
\textbf{Encryption management} anonymises or creates a hash  for each CID generated by IPFS. The anonymisation is used for protecting CIDs from unauthorised access. The layer also keeps CIDs and their hashed versions in a local database.    
\\
\textbf{Blockchain virtual machine and smart contracts factory} hosts the following smart contracts for storing and monitoring immunity passports through a Blockchain. 
\begin{itemize}
\item \textbf{Policy contract} involves two functions, called as \texttt{purpose()} and \texttt{vote()}. The former records what operation (i.e., read, write etc.) will be executed by which actor on citizen's data. Each record shows a purpose of data processing by the actor who is a third party processing passports' data. As an example, an actor can be the provider of a cloud-based service who has a contract with medical centers in order to collect and profile the data for analytic purposes.

The latter function retrieves the purposes of data processing from the Blockchain and stores the positive/negative  citizen's consent for each retrieved purpose in the Blockchain. The deployer of the contract is administrator that provides citizens with deployment address to receive their votes (positive or negative consents).
\item \textbf{Log contract} sends the anonymised version of CID along with its creation time into the Blockchain network. Such records are used as public keys for accessing the passports details. The \emph{contract deployers} are trusted administrators identified by medical centres. 
\item \textbf{Access contract} logs every access to CIDs in a Blockchain. It logs the operation (i.e., read, write, delete, and so on) which is executed by an actor on citizens' data within IPFS and submits it to the Blockchain. The contract is deployed by the \emph{access control manager} in the system. 
\item \textbf{Verification contract} provides the audit trail of actors processing or accessing to citizens' data. The contract involves a function, called as \texttt{verify} which identifies the actors collecting or manipulating vaccine records without getting positive consents from citizens as violators. The function is activated by a trusted third party, as referred to \textit{arbiter}. 
\end{itemize}
The deployers or agents existing in upper or lower layers  interact with the proposed contracts to record data in Blockchains.  
\\
\textbf{Access control and policy management} establishes a role-based mechanism for reading or updating citizens data. Users based on their roles can access to CIDs and vaccine passports details, and a copy of such access will be sent into Blockchain. 

The layer also determines a set of privacy policies in forms of $\langle ``actor", ``operation", ``purpose"\rangle$, as referred \textit{data processing purposes}. An administrator in the layer communicates with the smart contract factory to store such purposes of data usage in a Blockchain.
\\
\textbf{User interface} implements a user-friendly and front-end decentralized application (DApp) for citizens in order to readily interact with the platform. Technically, it is connected to the contracts' interfaces created and hosted on Ethereum Blockchain virtual machine. The interface also enables citizens to retrieve the purposes of data processing from Blockchain with a more legible format and get their votes (positive/ negative consent) to the predefined purposes. In fact, the citizens' consents will be considered as the inputs for \texttt{vote()} function involved in the \textit{policy} smart contract.

There are four phases for realizing the architecture: \textit{agreement}, \textit{passport creation}, \textit{access control} and \text{verification}.

\begin{figure}[t!]
\center
\includegraphics[width=0.45\textwidth]{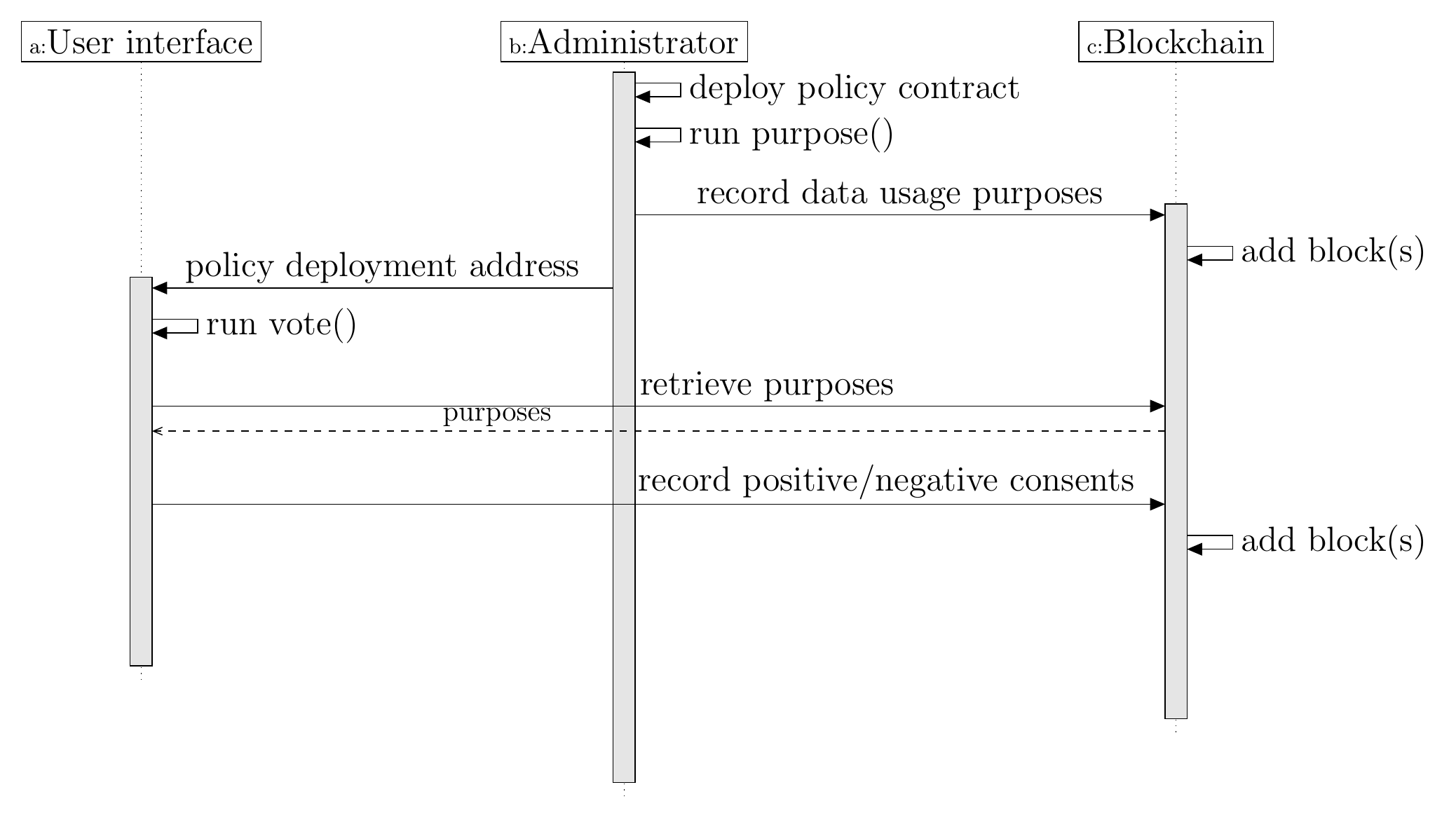}
\caption{A protocol for agreement phase}
\label{fig:agree}
\end{figure}

\subsection{Agreement phase}

This phase presents a protocol for demonstrating the attractions among citizens, system administrator and Blockchain for recording purposes of data processing and citizens' consents. Figure~\ref{fig:agree} shows the protocol in the form of a sequence diagram. As seen, the main entities are user interface, system administrator and Blockchain. The administrator as the data controller, first, deploys the policy contract and activates the purpose function to send data usage purposes into Blockchain. Precisely, each record contains: (i) \textit{actor} who will update or collect citizen' passport data (ii) \textit{operation} that shows what actions (e.g., read, write etc.) will be carried out by the actor on citizens' data, and (iii) \textit{purpose} that describes the operation is used for what. Once such records have been added to the Blockchain network, the administrator provides a user interface with the deployment address of policy contract in order to make the records accessible to end users (citizens). The user interface, then, by activating the vote function, enables users to retrieve the data usage purposes from Blockchain and freely give their consents to them before any processing on their personal data. The users' votes will be kept in the Blockchain as evidence for future verification. This phase realises Recitals (32) and
(43) of GDPR under which data subjects (citizens) should 
give their consent for any operation undertaken by data processors (actors) on their personal data.

\begin{figure}[t!]
\center
\includegraphics[width=0.45\textwidth]{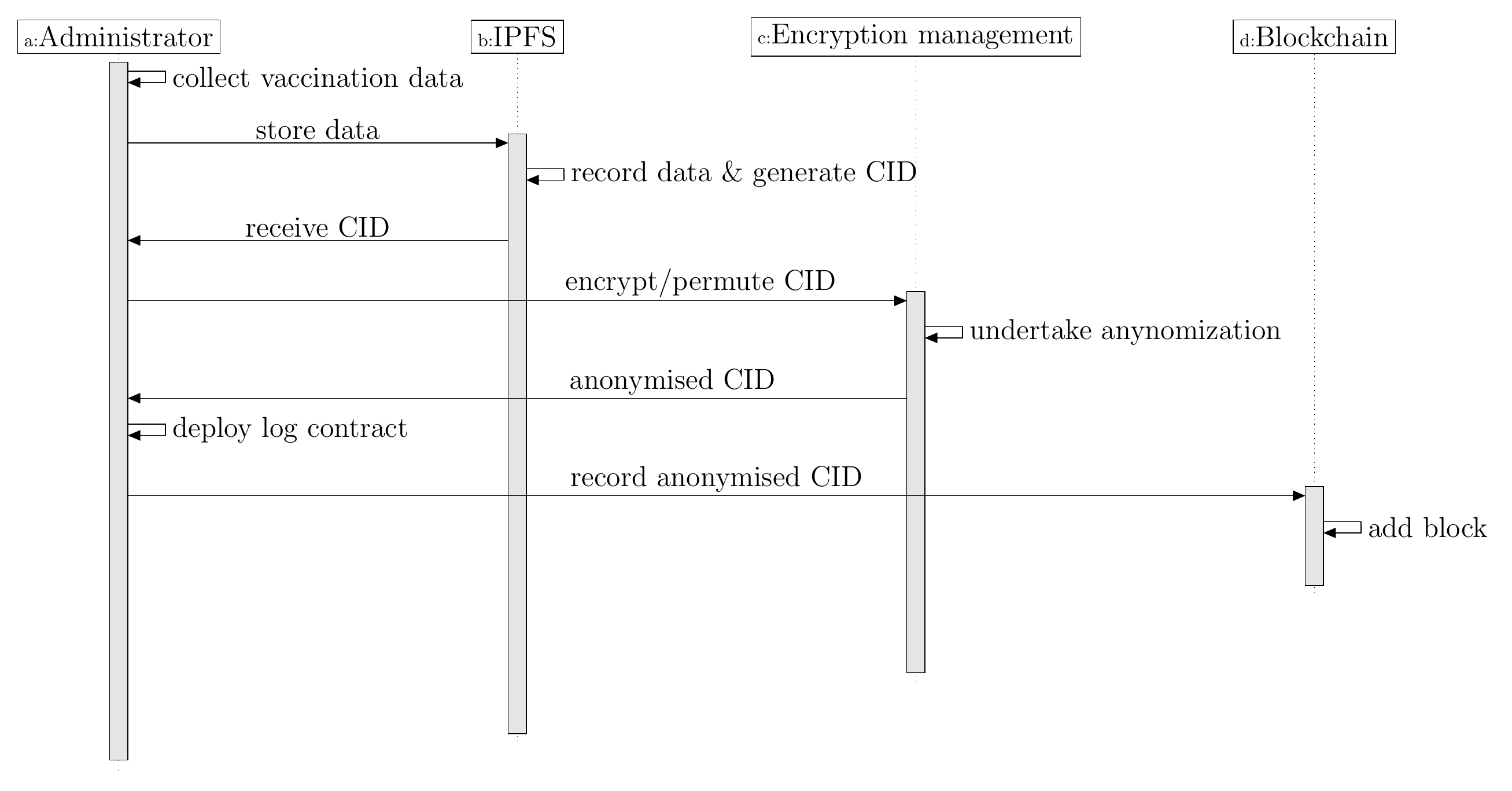}
\caption{A protocol for passport creation phase}
\label{fig:pass}
\end{figure}

\subsection{Passport creation phase}

This phase represents the steps in which the vaccine passports details will be stored in IPFS and their associated anonymised CIDs are recorded in Blockchain. Figure~\ref{fig:pass} depicts the protocol of the phase. After collecting citizens data by passport administrator, the data is sent to IPFS. Then, the data is recorded and a CID is automatically generated via IPFS. Once the CID has been received, it is forwarded to the encryption management layer so as to be anonymised. Following that, the passport administrator, by deploying log contract, submits the anonymised version of CID to Blockchain. 

There are several techniques for data anonymisation~\cite{Sedayao:2012} such as hashing, permutation among others, each of which can be exploited for mapping the CIDs into the anonymised ones. 

\begin{figure}[t!]
\center
\includegraphics[width=0.45\textwidth]{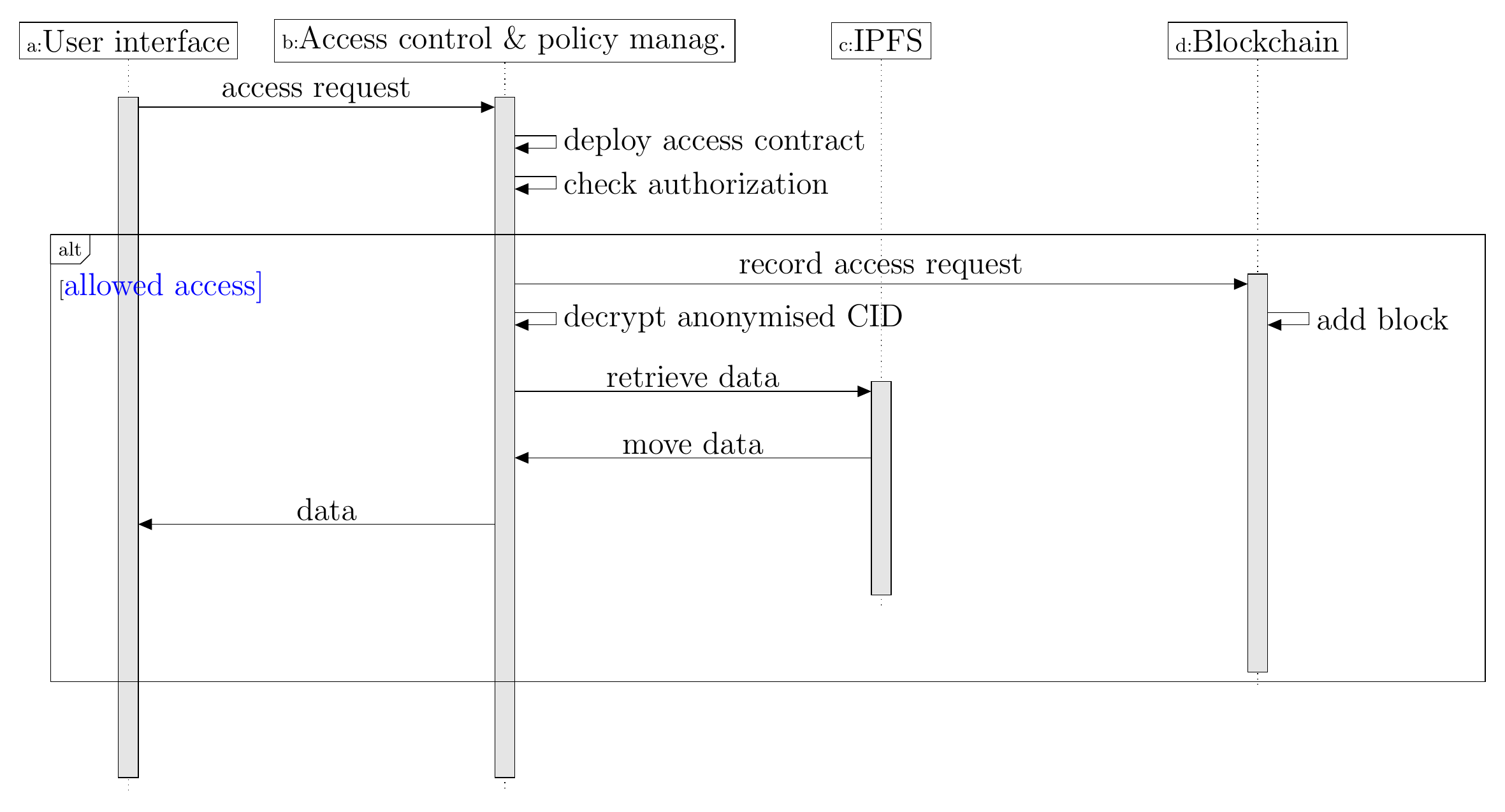}
\caption{A protocol for accessing to passport data}
\label{fig:access}
\end{figure}

\subsection{Access control phase}
This phase monitors and verifies all the access requests to passport citizen data. The sequence diagram depicted in Fig.~\ref{fig:access} is a protocol for the access control. Citizens and trusted parties through user interface are able to send their request for access to passports data. The request also contains  the operation (such as update and so on) that will be carried out on the data. Upon the receipt of request, the access control management service's agent checks the authorisation of requester. In case of authorised access, the agent  deploys the access smart contract in order to record \textit{requester ID}, \textit{access time} and \textit{permitted/ executable operation(s)} (e.g., view, update etc.) in the Blockchain.\footnote{For privacy purposes, a hashed version of requester ID, which refers to their Blockchain account is stored on-chain.} Such records are used for future verification. Following that, the hashed/anonymised version of CID is decrypted, and finally the passport data is retrieved from IPFS to be accessible for the requester.    

\begin{figure}[t!]
\center
\includegraphics[width=0.45\textwidth]{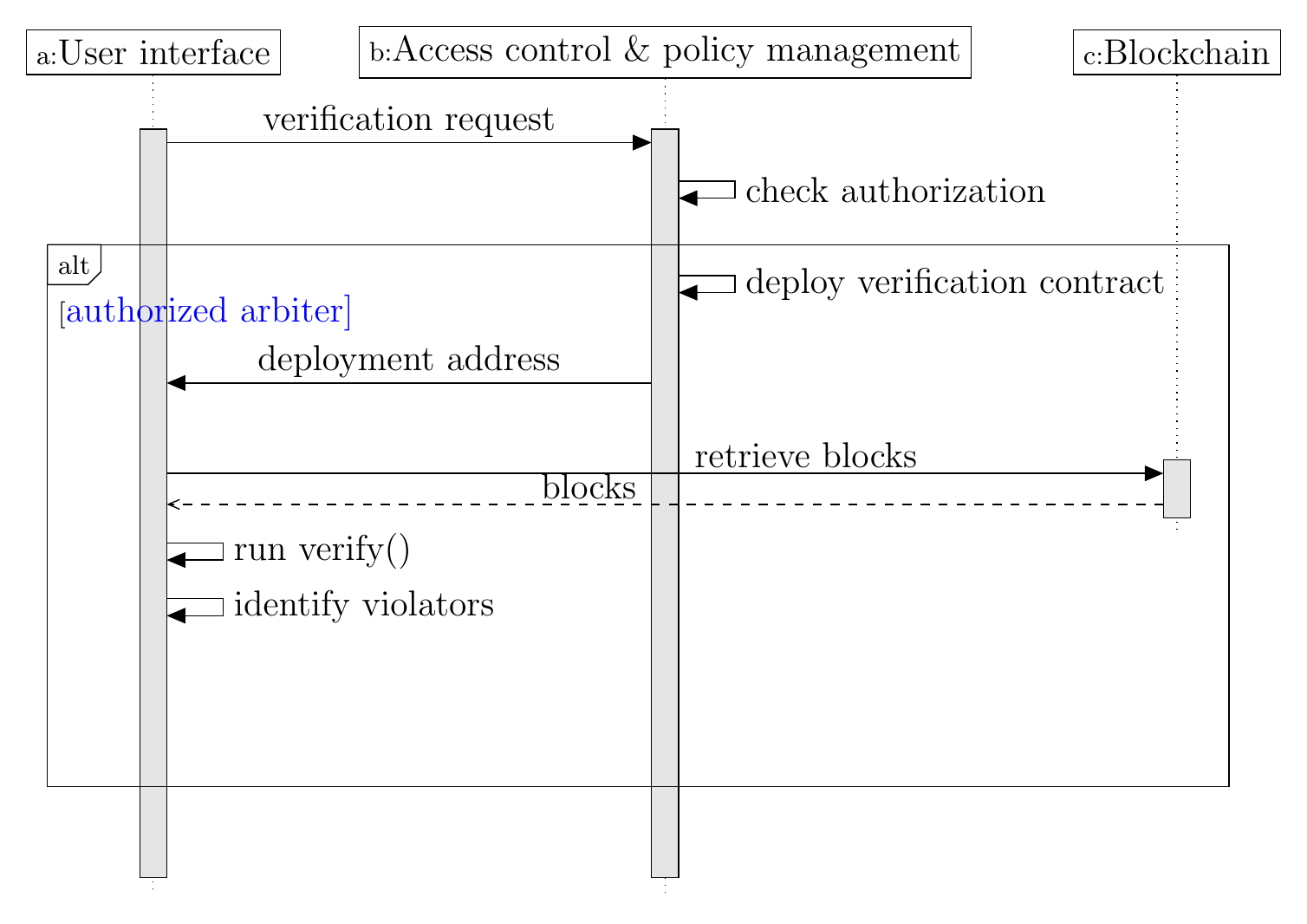}
\caption{A protocol for verifying GDPR violations}
\label{fig:verify}
\end{figure}

\subsection{Verification phase}

This phase, through the protocol represented in Fig.~\ref{fig:verify}, detects the violators who access or manipulate citizens data without getting their consents. 
The arbiter via user interface sends their verification request for reporting violators once claimed by citizens or legal offices. After approving the authorization of arbiter by access control management service's agent, the verification contract is deployed and its address is forwarded to the arbiter. The deployment address enables the arbiter to access the records already stored by both \textit{policy} and \textit{access} contracts. Then, the arbiter, by running the verify function, flags any violation on citizens data and identifies violators. The verify function checks the following items to detect violations:

\begin{enumerate} 
\item whether the  actors stored by access contract conform to those logged via policy contract or not;
\item whether the operations of each actor recorded through access contract conform to those recorded via policy contract or not;
\item whether the operations of each actor logged by access contract were already confirmed by the data subject (citizen) or not.
\end{enumerate}

Assuming that $A_c$ is the set of actors with  positive consent from data subject via the policy contract; $A_e$ is the set of actors executed operations on citizen passport's data and recorded by the access contract; $O_{a}$ is the set of operations of actor $a \in A_c$ got positive consent from data subject via the user policy contract; and  
$\mathcal{O}_{a}$ is the set of operations executed by $a  \in A_c$  on passport data and recorded via the access contract.

Given these assumptions, Algorithm~\ref{verify1} presents the verification of actors implemented as a part of \textit{verify} function.

\small
   \begin{algorithm}
    \caption{ ~Verifying actors}\label{verify1}
\hspace*{\algorithmicindent} Let $V$ be a set denoting violators \\
   \hspace*{\algorithmicindent} \textbf{Input:} policy \& access deployment addresses\\
    \hspace*{\algorithmicindent} \textbf{Output:} $V$
    \begin{algorithmic}[l]
    \Function{verify}{}
\State $ V \gets\emptyset ;$
    \If {$A_e \not \subseteq A_c$} 
 \State $ V \gets V \cup A_e\!\setminus \!A_c;$
     \EndIf
     \For {$\mbox{all}~a \in A_c$}
      \If {$\mathcal{O}_{a} \not \subseteq O_{a}$} 
   \State $ V \gets V \cup \{a\};$
    \EndIf
\EndFor
    \State $\textbf{return}~~V;$
    \EndFunction
    \end{algorithmic}
    \end{algorithm}
\normalsize
    
As seen from the algorithm, a violation is flagged if: (i) an actor processes passport data without the confirmation of data subject; and (ii) an accepted actor executes an operation already rejected by the data subject.

\begin{figure}[t!]
\center
\includegraphics[width=0.45\textwidth]{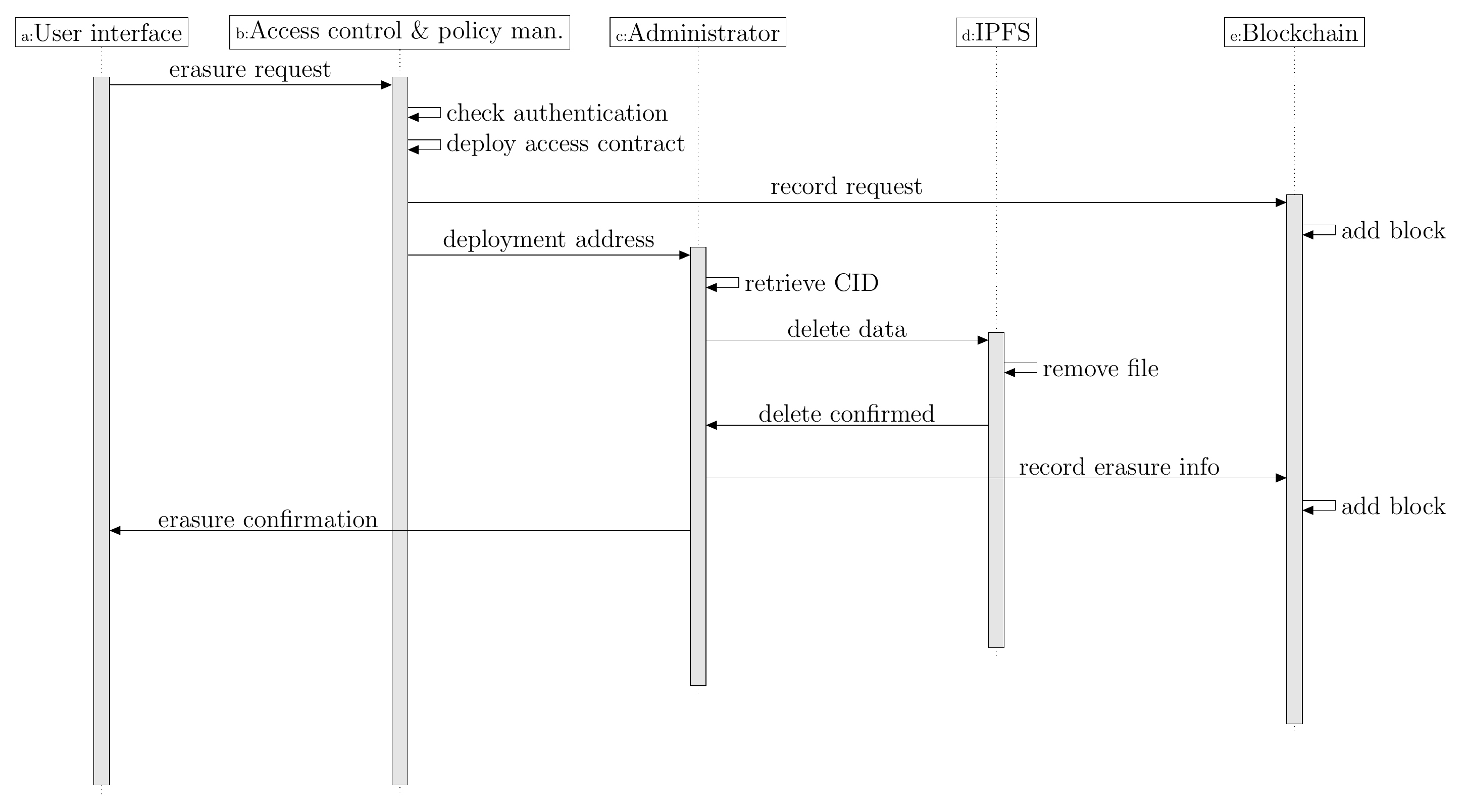}
\caption{A protocol for erasure of vaccine data}
\label{fig:rightForg}
\end{figure}

\subsection{Right to be forgotten}

Citizens have the right to get from the passport administrator (data controller) the erasure of their data without any delay (Art.~17 of GDPR). In order to realize this GDPR principle through our proposed architecture, a protocol is presented in Fig.~\ref{fig:rightForg}. It provides a basis for data accountability of administrators to track whether the citizen's data have been deleted without undue delay. As represented from the figure, a citizen, through user interface, is able to submit a data erasure request, which is received by access control management service's agent. The identification and permission of the citizen is then verified by the agent and the access smart contract is deployed so as to record a copy of the request in a Blockchain. Upon the receipt of the deployment address of the contract by the passport administrator, the CID related to vaccine passport's data is collected and the citizens' personal data is removed within IPFS. After erasing the data, a confirmation is sent to the citizen. Moreover, a copy of such confirmation denoting the data have been removed by who and when is stored in the Blockchain as evidence for future verification.     

In order to detect any violation with, the \textit{verification} smart contract is extended to include a \texttt{erase\_verify()} function. By retrieving the blocks containing erasure requests/ confirmation and created by the access contract, the function flags the violator if:
\begin{itemize}
 \item  the erasure confirmation has not been recorded by the administrator in the Blockchain; or
    \item the time difference between erasure request and erasure confirmation logged by the administrator is greater than a short deadline already determined through the purposes of data processing in the agreement phase.
\end{itemize}

Both cases are investigated by the arbiter with regards to a claim received from the citizen who is the owner of the passport. For instance, after the submission of an erasure request, if the citizen observes that their data is still available in IPFS while the deadline had been passed, a claim can be made by the citizen and submitted to the arbiter. The claim should involve the erasure request's time and anonymised CID.

\section{Experimental Results}
\label{expr}
Our experiments cover the evaluations related to the creation of vaccine passports using IPFS and the implementation of our proposed smart contracts using Blockchain test networks. 

\subsection{IPFS CID Generation Time for Vaccination Passports}
To demonstrate the scalability of our proposed solution, the CID generation time in IPFS was evaluated. We chose to measure the time taken to generate 10, 20... up to 100 CID values for simulated vaccination passports which may be added to the IPFS network. Our vaccination passport is encoded as a JSON object, and the contents are derived from the fields used by NHS Scotland in real-world vaccination scenarios (described in Section \ref{sec:arch}). The JSON string for an example passport is shown in Appendix \ref{appendix:passportjson}.

A script was used to generate the passport data. Each passport created will generate a unique Community Health Index (CHI) number: a 10-digit value used to identify patients in Scotland. The use of a unique CHI number for each passport ensures the IPFS CID generated will also be unique (recall that the IPFS CID is derived from the content of a resource using cryptographic hashing). Each passport object generated is 452 bytes in size.

\begin{figure}[h!]
\begin{center}
\includegraphics[width=0.75\columnwidth]{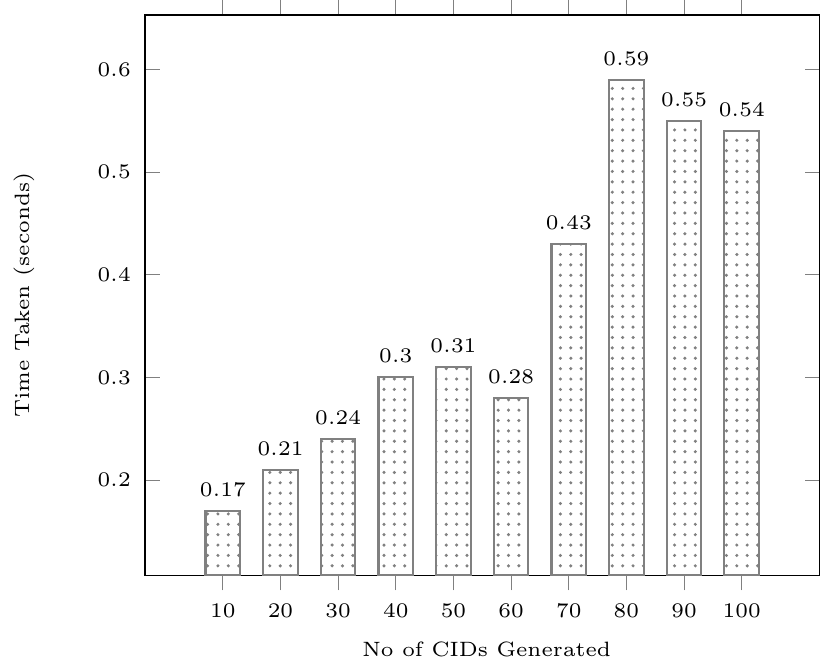}
\label{dag01}
\caption{IPFS CID Generation Time}\label{fig:ipfs_generation_time}
\end{center}
\end{figure}

A private instance of the IPFS network \cite{IPFS2021b} was created as a Docker image for the purposes of this evaluation. A private IPFS network (one which is isolated from the public network) was used to ensure no experimental data was accidentally added to the public P2P network. The Linux \textbf{time} \cite{LinuxTime} command was used to monitor the time taken for the IPFS command to generate CIDs for 10, 20, ... 100 CIDs. An example of the $time$ command used in conjunction with IPFS for evaluating time taken for generating ten unique vaccination passport CIDs is as follows\footnote{We attribute this approach to redirecting the time output to \cite{stackoverflow}}:

\begin{lstlisting}
{ time ipfs add -r 10 ; } 2> 10_result.txt
\end{lstlisting}

In the above code listing, we monitor the time taken (user+sys time) for the \textbf{ipfs add -r 10} command to execute. The \textbf{ipfs add -r} command is used to generate CIDs recursively for all content (i.e. 10 simulated vaccination passport JSON objects in this example) within a folder named \textbf{10}. The same measuring approach is taken for 20 unique passports, 30 and so on.

Figure \ref{fig:ipfs_generation_time} shows the generation time for 10, 20... 100 CIDs derived from simulated vaccination passport data. For 10 CIDs, the time varied between 0.37s and 0.59s. On 6 June 2021 there were 387,286 total vaccinations given to UK citizens based on figures released by the UK government \cite{UK2021}. If we assume a vaccination passport entry was to be created for all 387,286 vaccination events, and assume it takes a maximum of 0.59s to generate 100 CIDs, our results show that it would take around \textbf{38 minutes} to generate a CID for all passports. This demonstrates that the CIDs can scale to a large population size.

\subsection{Investigation of proposed smart contracts}

A prototype has been developed using both Ganache~\cite{ganache:2021} and Ropsten~\cite{Ropsten:2021} test networks. We implemented our smart contracts on Ethereum via Solidity language.  Ganache is a local test network that provides multiple default gas and ether values, which can be applied as a currency to change Blockchain states when running function calls. Ropsten is a public test network involving a set of miners and gives detailed information relevant to miners. However, it has a gas limit of 4712388 for executing a smart contract. Our proposed smart contracts have been written with a minimum gas usage for each function activation. They were compiled and successfully tested using Remix, being a browser-based development environment for Solidity. The contracts \textit{Policy}, \textit{Log},  \textit{Access} and \textit{Verification} were deployed in the aforementioned networks.  The amount of gas used for contract deployment was 792065 for \textit{Policy}, 157339 for \textit{Log}, 796253 for \textit{Access}, and 1223998 for   \textit{Verification}. The results represent the computational  cost  for executing each contract. However, changing the number of actors and their access requests has an impact on the transaction costs and mining time. 


\begin{figure}[t!]
\center
\includegraphics[width=0.7\columnwidth]{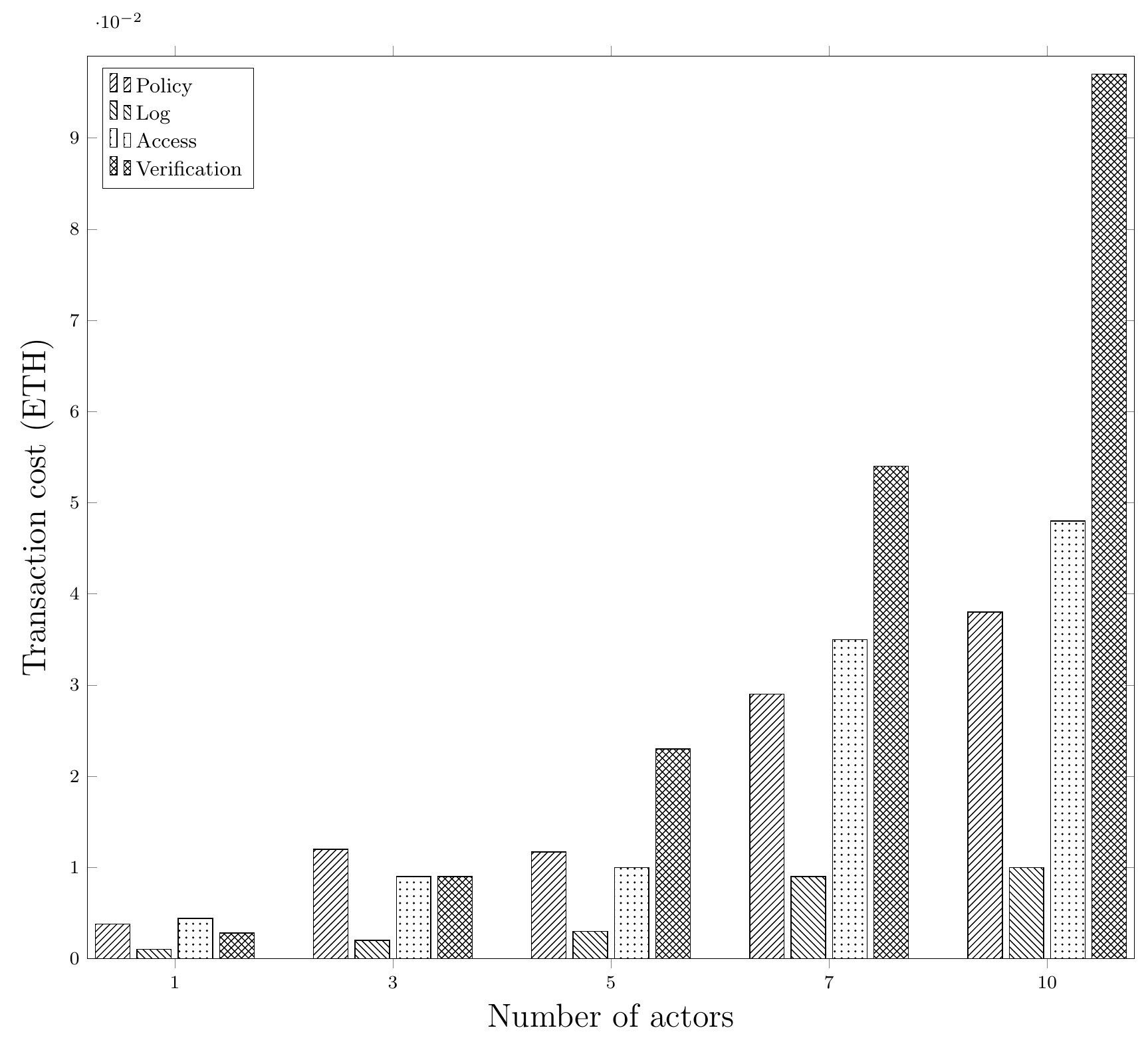}
\caption{The relationship between number of actors and  cost}
\label{fig:expgas}
\end{figure}

\subsubsection{Number of actors and transaction costs}

The experiment which changes the number of actors, ranging from one to ten, evaluates the cost required for activating transactions. The assumption is that the gas price unit is  20 \textit{gwei} and the number of operations (i.e., read, write, delete) carried out by each actor on vaccine data is three. Our proposed smart contracts have been deployed in the Ganache test network. We calculated the average costs in Ether (ETH) after five times execution of functions with different parameters (values). Figure~\ref{fig:expgas} illustrates the results of this experiment. As seen, the lowest costs in Ether are allocated to the transactions with one actor and the highest values belong to those involving ten actors.\footnote{The actors for \textit{Log} contract are citizens, whose anonymised CIDs are recorded in Blockchain.} Furthermore, when the number of actors increases, the verification cost rises more sharply compared to the other contracts. In fact, the  \textit{Verification} contract has a high complexity, since it must call both \textit{Policy} and \textit{Access} smart contracts in order to check the GDPR compliance of actors and report violators.   

\begin{figure}[t!]
\center
\includegraphics[width=0.7\columnwidth]{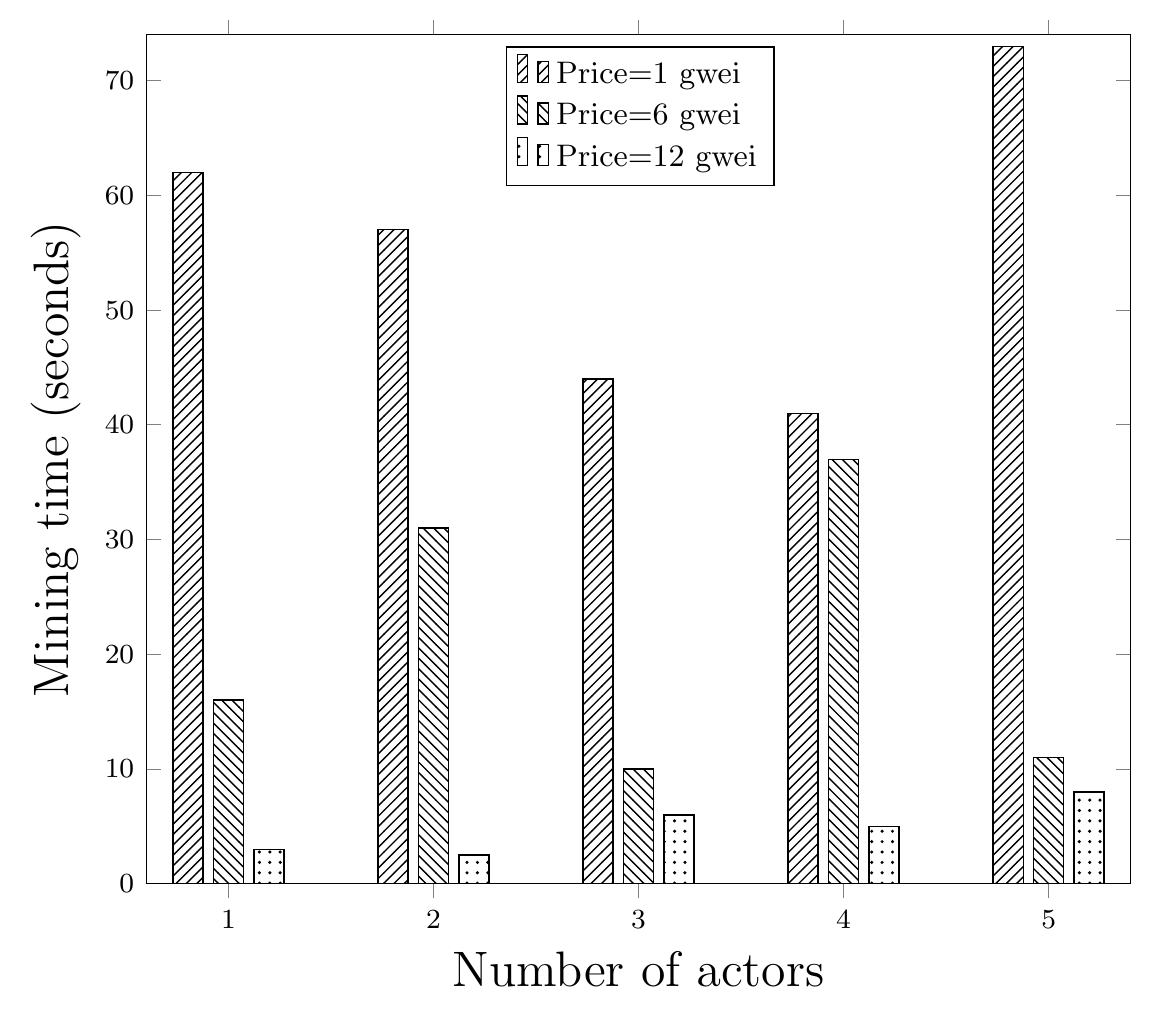}
\caption{The relationship between number of actors and mining time}
\label{fig:exptime}
\end{figure}

\subsubsection{Number of actors and mining time}
This investigation represents the impact of changing the number of actors on the time taken for mining process under different gas prices. The number of actors is varied from one to five and each of which executes two operations on citizens data. We have various scales of gas prices (i.e., 1, 6 and 12 \textit{gwei}). The Ropsten test network was used to get the results of this experiment, as it gives a measurement of the time taken from execution to mining of a block. The \textit{verification} contract was deployed in Ropsten and its \textit{verify} function was  activated five times to reach an average mining time. Figure~\ref{fig:exptime} shows the results of this evaluation. Given a fixed gas price, there is a fluctuation in the trend of the chart. As a result, the time totally depends on the interest of miners to validate/ mine the  transactions created from the activation of \textit{verify} function and the number of actors does not have an impact on mining time. However, when the amount of gas price increases, the mining time decreases significantly. It shows that higher gas prices motivate miners to accelerate the mining process and block creation.  

\section{Conclusion}
\label{conclu}

The design of a Blockchain-based platform for the creation of online COVID-19 vaccine certificates is proposed. The platform uses IPFS and smart contracts to support privacy of citizens' information and supports data accountability for third party access/ processing of this information. The purposes of data processing (carried out by actors) is automatically sent into a Blockchain and the platform enables citizens to give a vote (positive /negative consent) for each purpose via a smart contract. The proposed approach meets GDPR requirements, and only non-sensitive data is stored within the Blockchain for auditing purposes. Compared to other Blockchain-based platforms for cloud and IoT ecosystems for keeping and verifying patient data~\cite{BaratiRana:2020,BRPT:2020}, our platform  provides a technical solution for checking data erasure requests, which is a significant user right in GDPR (referred to as ``right to be forgotten"). A vaccine passport template has been implemented as a prototype, and our evaluations shows that it takes less than one second to generate 100 passport CIDs in IPFS. The created smart contracts have been tested in both Ganache (on a local machine) and Ropsten (a global Blockchain network) environments and results in the transactions cost increasing noticeable when the number of actors increases (as expected). The results of these experiments can be used to support capacity planning of a vaccine certificate network. Given a fixed gas price used for execution of smart contract opcodes, the investigation demonstrates that miners can take an arbitrary time for mining blocks. Future work focuses on the implementation of both access control and encryption management layers of the designed architecture. The development of the proposed platform in cloud environment and the management of CIDs generated by IPFS remain other aspects for future investigation. \\

\noindent {\bf Acknowledgment:} This work has been carried out in the GLASS (SinGLe Sign-on eGovernAnce paradigm based on a distributed file exchange network for Security, transparency, cost effectiveness and truSt) project \cite{glass2021}.

\bibliographystyle{IEEEtran}
\bibliography{main}

\appendix
\noindent {\bf Example Vaccination Passport JSON object}: 
\label{appendix:passportjson}
Example of JSON encoded vaccine passport data for evaluation of IPFS CID generation time --  based on data used by NHS Scotland. A total of 100 vaccination passport objects were created of 452 bytes each. A unique number was provided in the CHI field (as unique patient identifier) when creating a passport to ensure uniqueness of hash value when generating the IPFS CID.

\small
\begin{lstlisting}
{
"COVID-19 Vaccination Status": {
  "CHI": "0000000001",
  "Surname(s)": "Doe",
  "Forename(s)": "John",
  "DOB": "02/01/1965",
  "Disease targeted": "COVID-19",
  "Country of vaccination": "Scotland",
  "Issued by": "NHS Scotland",
  "Doses received": "2",
  "Dose 1 of 2": "01/06/2021",
  "Manufacturer": "Moderna Biotech Spain S.L.",
  "Vaccine medicinal product": 
              "COVID-19 Vaccine Moderna",
  "Vaccine/Prophylaxis": 
               "SARS-CoV-2 mRNA vaccine",
  "Dose 2 of 2": "30/06/2021"
  }
}
\end{lstlisting}
\normalsize

\end{document}